\documentclass[aps,prb,twocolumn,showpacs]{revtex4-1}
\usepackage{graphicx}
\usepackage{graphics}
\usepackage{amsmath}
\usepackage{amsfonts}
\usepackage{amssymb}
\usepackage{epstopdf}
\usepackage{makeidx}
\usepackage{epsfig}
\usepackage{bm}
\usepackage{color}

\usepackage[unicode=true, bookmarks=false,breaklinks=false,pdfborder={0 0 1},backref=false,colorlinks=false]{hyperref}
\hypersetup{hypertex}
\usepackage{breakurl}
\hypersetup{backref,colorlinks=true,citecolor=blue,linkcolor=blue,filecolor=blue,urlcolor=blue}

\graphicspath{{draft_figures/}}

\begin{document}

\newcommand{\be}   {\begin{equation}}
\newcommand{\ee}   {\end{equation}}
\newcommand{\ba}   {\begin{eqnarray}}
\newcommand{\ea}   {\end{eqnarray}}
\newcommand{\ve}   {\varepsilon}

\newcommand{\CAIO}[1]{\textcolor{red}{\fbox{Caio} {\sl#1}}}
\newcommand{\JESUS}[1]{\textcolor{green}{\fbox{Jesus} {\sl#1}}}
\newcommand{\RICARDO}[1]{\textcolor{blue}{\fbox{Ricardo} {\sl#1}}}
\title{Edge magnetization and local density of states in chiral nanoribbons}

\author{A. R. Carvalho, J. H. Warnes, and C. H. Lewenkopf}
\affiliation{Instituto de F\'{\i}sica,
Universidade Federal Fluminense, 24210-346 Niter\'oi, RJ, Brazil}

\date{\today}

\begin{abstract}
We study the edge magnetization and the local density of states of chiral graphene nanoribbons 
using a $\pi$-orbital Hubbard model in the mean-field approximation. 
We show that the inclusion of a realistic next-nearest hopping term in the tight-binding Hamiltonian 
changes the graphene nanoribbons band structure significantly and affects its magnetic properties.
We study the behavior of the edge magnetization upon departing from half filling as a function of the 
nanoribbon chirality and width. We find that the edge magnetization depends very weakly in the 
nanoribbon width, regardless of chirality as long as the ribbon is sufficiently wide. We compare 
our results to recent scanning tunneling microscopy experiments reporting signatures of magnetic 
ordering in chiral nanoribbons and provide an interpretation for the observed peaks in the local 
density of states, that does not
depend on the antiferromagnetic interedge interaction.
\end{abstract}

%

\maketitle

\section{Introduction}
\label{sec:introduction}

Graphene shows a large variety of novel and unique electronic properties 
\cite{CastroNeto2009}. Of particular interest is the emergence of magnetism in 
graphene nanostructures \cite{Yazyev2010}. The prediction of localized states at the 
edges of graphene nanostructures \cite{Nakada1996}, that are believed to give rise to 
edge magnetization \cite{Fujita1996,Kusakabe2003,FernandezRossier2007} with 
potential applications in spintronics, \cite{Son2006Nature} has attracted a lot of 
theoretical and experimental attention. 
Electronic structure calculations indicate that graphene nanoribbons (GNRs), 
depending on the crystallographic orientation of their edges, exhibit a ferromagnetic 
spin alignment along the edges and an antiferromagnetic interedge ordering 
\cite{Fujita1996,Son2006Nature,Pisani2007,Jung2009PRL,Yazyev2011}. 
Several experiments report evidence of edge states 
\cite{Klusek00,Kobayashi05,Ritter09,Tao2011}, but direct observations of edge 
magnetization in graphene remains rather elusive. 

The synthesis of GNRs was pioneered by lithographic patterning 
\cite{Chen2007,Han2007,Han2010}. 
This technique produces rough edges, that give rise to short-range scattering, 
detrimental to the electronic mobility  \cite{Mucciolo2009} and to the 
formation of local magnetic moments \cite{Wimmer2008}. More recently, by 
chemically unzipping carbon nanotubes, it became possible to obtain GNRs with 
very smooth edges \cite{Li2008,Jiao2010,Kosynkin2009}. In general, the latter are 
chiral, that is, their edges do not follow neither the zigzag nor the armchair 
high-symmetry orientations. The local density of states (LDOS) of ultrasmooth 
edge chiral GNRs was recently investigated using scanning tunneling microscopy/scanning tunneling spectroscopy 
(STM/STS) \cite{Tao2011}. The obtained STS spectra are the first direct experimental 
evidence of localized edge states in chiral GNRs. These results are the experimental 
motivation for this paper.

The theoretical studies of electronic properties of graphene nanoribbons with 
arbitrary edges date back to a pioneering paper in the field \cite{Nakada1996}.  
It has been established that, for sufficiently wide ribbons, there is always an 
enhancement of the density of states (DOS) due to dispersionless zero-energy
edge states, except for GNRs with armchair terminations \cite{Akhmerov2008}.
In chiral GNRs, as in the zigzag case, electron-electron interactions split the zero-energy bands 
and give rise to edge magnetization \cite{Yazyev2011,Sun2011}; whereas the 
Hubbard mean-field calculations \cite{Yazyev2011} indicate that local magnetization 
appears whenever the noninteracting DOS in enhanced \cite{Yazyev2011}, 
density functional theory (DFT) calculations point to a sharp suppression of the edge 
magnetization for chiralities close to the armchair orientation \cite{Sun2011}. 

In this paper, we investigate edge magnetization in graphene chiral nanoribbons. 
More specifically, we systematically study the local magnetization as a function of 
the GNRs chirality, width $W$, and doping, the latter cast in terms of a chemical 
potential $\mu$. 
We use a Hubbard mean-field model and include a next-nearest-neighbor (nnn) 
hopping term $t'$ in the tight-binding description. The latter is necessary to 
reproduce the low-energy DFT band structure calculations \cite{Wakabayashi2012}.
We find that, for sufficiently large $W$, the local magnetization is a function of the 
chirality with a negligible dependence of $W$. The study of the GNRs edge magnetization 
$M$ as a function of the chemical potential $\mu$ reveals a strong 
correlation between $M$ and some characteristic features of the band structure.
We compare our calculations of both the DOS and the DOS with recent (STS) experimental results recently obtained for chiral
GNRs \cite{Tao2011}.
Our results indicate that, by using realistic values of $t'$ in the tight-binding model, 
the simple interpretation reported in the literature \cite{Tao2011,Yazyev2011} of the 
experimentally observed peaks in STS spectra in terms of edge magnetic ordering is 
hardly justified. This conclusion calls for further experimental and theoretical investigations 
for evidence of magnetism in GNRs.

The paper is organized as follows. In Sec.~\ref{sec:theory}, we present the model 
Hamiltonian used in this study, introduce the notation to describe the geometry chiral 
edges, and review the theory. Our results are presented in Sec.~\ref{sec:results}. 
We begin by analyzing the edge magnetization for zigzag GNRs, that serves as a 
guide for the discussion that follows. Next, in Sec.~\ref{sec:chiral}, we study the 
edge magnetization in chiral GNRs as a function of chiral angle, width and doping. 
Finally, in Sec.~\ref{sec:comparison}, we show our results for the LDOS and discuss 
the connection between our findings and experimental results.  We present our conclusions 
in Sec.~\ref{sec:conclusions}.

\section{Theoretical model}
\label{sec:theory}

\subsection{Chiral nanoribbons: lattice parametrization}
\label{sec:chiral-ribbon}

\begin{figure}[h!]
\centering \includegraphics[width=0.9\columnwidth]{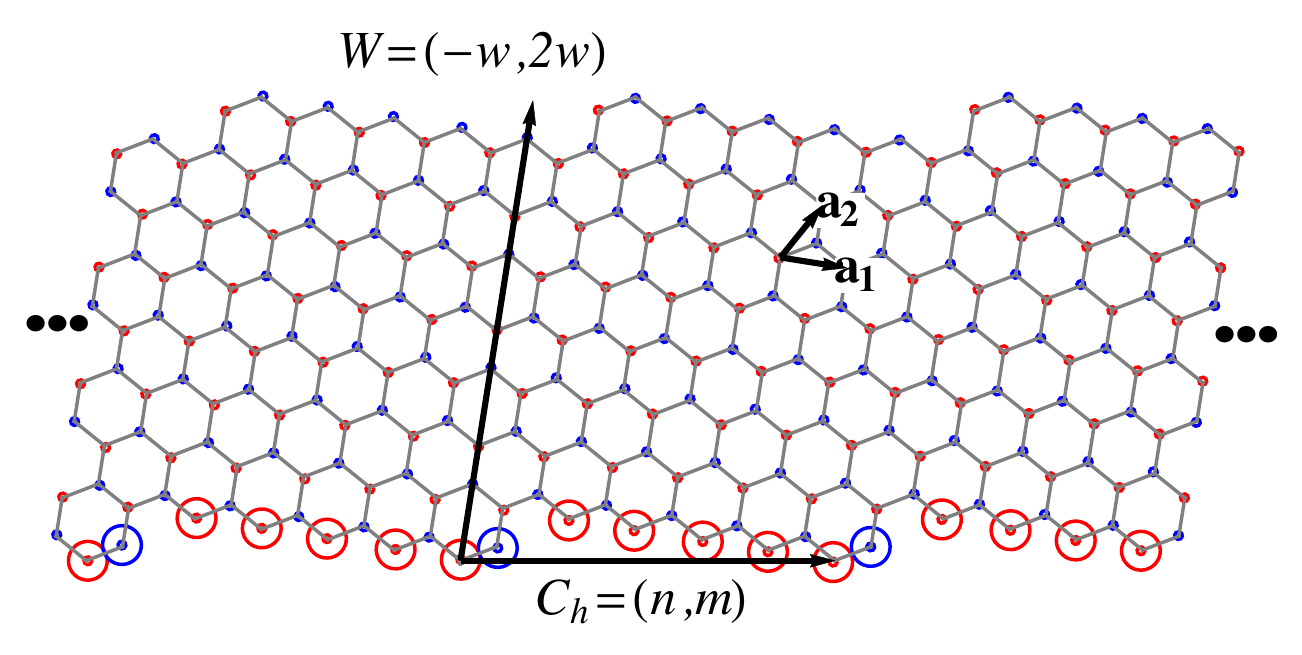} 
\caption{(Color online) Representation of a chiral nanoribbon. The chiral vector is 
${\bf C}_h = (5, 1)$, and the width is characterized by ${\bf W}= (-4, 8)$. The sites 
with dangling bonds at the GNR edges are indicated by circles.}
\label{fig:chiralGNR}
\end{figure}

We define the primitive unit cells (PUCs) of a chiral GNR in terms of their widths and
the crystallographic direction of their edges. \cite{Akhmerov2008} The GNR longitudinal 
orientation is characterized by the translation (or chiral) vector ${\bf C}_h$, see 
Fig.~\ref{fig:chiralGNR}, defined as
\be
\label{eq:chiral_vector}
{\bf C}_h = n{\bf a}_1+m{\bf a}_2 \equiv (n, m), 
\ee
where $n$ and $m$ are integers, whereas ${\bf a}_1$ and ${\bf a}_2$ are the 
lattice unit vectors. 
The length of the translation vector is $a= a_0 \sqrt{m^2 + mn + n^2}$, 
where $a_0\approx 0.246$ nm  is the graphene lattice constant. 
For later convenience, we write ${\bf C}_h$ in terms of its projection on the zigzag and armchair directions
\be
\label{eq:chiral_vector_projection}
{\bf C}_h = {\bf C}_{h,{\rm zz} }+ {\bf C}_{h,{\rm ac}} = (n-m){\bf a}_1 + m({\bf a}_1+{\bf a}_2) .
\ee

In general, ${\bf C}_h$ does not provide a precise characterization of the edges, 
since GNRs with the same  translation vector can have a different number of edge atoms 
$N_e$ and dangling bonds $N_d$ per unit cell. The constraint that neither $N_e$ nor $N_d$ 
can be smaller than $m+n$, \cite{Akhmerov2008} is used to define ``minimal edge" GNRs 
\cite{Akhmerov2008,Jaskolski2011} where $N_e=N_d=m+n$. In this case, ${\bf C}_h$ 
describes the nanoribbon edges unambiguously. For the sake of simplicity, in this paper, we 
consider only minimal edge chiral GNRs.

The GNR orientation is also often specified by the chiral angle $\theta_c$, defined as
\begin{equation}
\label{eq:theta_c}
\cos\theta_c= \frac{{\bf C}_h \cdot {\bf a}_1}{\lVert {\bf C}_h \rVert\lVert {\bf a}_1 \rVert}=
\frac{2n+m}{2\sqrt{n^2+m^2+nm}}.
\end{equation}

Due to the symmetry of the honeycomb lattice, $0^\circ \le \theta_c\le 30^\circ$ 
accounts for all possible crystallographic direction.
The high-symmetry cases are those where the GNR edges correspond to the 
zigzag and armchair directions, that is, $\theta_c=0^\circ$ with a translation vector 
$(n,0)$ and $\theta_c=30^\circ$ with $(n,n)$, respectively. 
GNRs, whose edges are neither armchair nor zigzag, are called chiral. 

The GNR width is conveniently characterized by \cite{Yazyev2011}
\begin{equation}
{\bf W}=-w{\bf a}_1+2w{\bf a}_2 \equiv(-w,2w),
\end{equation}
where $w$ is an integer. The vector ${\bf W}$ is parallel to the armchair 
lattice orientation, see Fig.~\ref{fig:chiralGNR}. The width of the chiral (and zigzag) 
GNRs is $W=\sqrt{3}wa_0\cos\theta_c$.

\subsection{Model Hamiltonian: Electronic structure }
\label{sec:Hubbard-mean-field}

The ground-state magnetic ordering driven by electron-electron interaction  has 
been extensively studied for GNRs with zigzag edges by a number of methods 
\cite{Wakabayashi2012}.
Band structures calculated by density functional theory with local spin-density 
approximation (DFT-LSDA) \cite{Son2006Nature,Son2006PRL} show remarkable 
agreement with those obtained from a tight-binding model with a Hubbard term in 
the mean-field approximation \cite{Hancock2010}. Further studies
treating the $e$-$e$ interaction beyond mean-field, such as Hartree-Fock with 
configuration interactions \cite{Dutta08,Dutta2012} and quantum Monte 
Carlo \cite{Feldner2010,Golor2013a,Golor2013b}, confirm that the mean-field 
approximation provides a good description of the magnetic ground-state properties of zigzag GNRs.

In this study, we use the Hubbard mean-field approximation to compute the electronic 
and magnetic properties of GNRs. As discussed, this simple model leads to 
results that agree with more sophisticated methods. Moreover, it allows for assessing the 
ground-states properties of GNRs with large primitive unit cells at a very modest 
computational cost. The model Hamiltonian reads
\begin{align}
\label{eq:Hamiltonian}
H =& - t\sum_{\langle i,j \rangle,\sigma} \!\!\left(a^\dagger_{i,\sigma}a^{}_{j,\sigma} 
+ \mbox{H.c.}\right) - t'\sum_{\langle\langle i,j \rangle\rangle,\sigma}
  \!\!\!\left(a^\dagger_{i,\sigma}a^{}_{j,\sigma} + \mbox{H.c.}\right)
 \nonumber \\
 & + U \sum_{i,\sigma}  n_{i,\sigma} \langle n_{i,-\sigma} \rangle,
\end{align}
where $a^\dagger_{i,\sigma}$ and $a^{}_{i,\sigma}$ , respectively, are the creation and annihilation operators of  electrons with spin projection $\sigma$ at site $i$, while
$n_{i,\sigma}=a^\dagger_{i,\sigma}a^{}_{i,\sigma}$ is the number operator and 
$\langle n_{i,\sigma}\rangle$ is its expectation value. The symbols $\langle \cdots \rangle$ 
and $\langle\langle \cdots \rangle\rangle$ indicate that the sums run over nearest-neighbor 
and next-nearest-neighbor lattice sites, respectively.

The magnitude of the on-site Coulomb energy $U$ in graphene systems is under 
current debate in the literature \cite{Schuler2013}. 
At the tight-binding level, several authors study parametrizations containing 
higher nearest neighbor contributions (see, for instance, Ref.~\onlinecite{Jung2013}) 
and their effect on breaking the particle-hole symmetry.
We take the pragmatic approach of taking the model parameters $t,$ $t',$ and $U$ that 
reproduce with great accuracy the  low-energy band structure and the local magnetization 
obtained by DFT-LSDA calculations for narrow nanoribbons \cite{Hancock2010}. 
In our study, we do not consider edge reconstructions \cite{Koskinen2008} and assume 
that the dangling bonds of undercoordinated edge atoms are passivated by hydrogen atoms, 
that have a similar electronegativity to the carbon ones. 

We use  the system translational invariance to write the eigenvalue problem 
in $k$-space. This is conveniently performed by means of the transformation
\begin{equation}
\label{eq:aml}
 b_{kl,\sigma} = \frac{1}{\sqrt{M}}\sum_m e^{i m k a} a_{ml,\sigma}
\end{equation} 
where $k$ is the wave number and $a$ is the length of the translation vector.
The sites are labeled $i=(m,l)$, where $l$ denotes the lattice site within the 
GNR primitive unit cell and $m$ labels the PUCs. 

The Hamiltonian Eq.~\eqref{eq:Hamiltonian} now reads
\begin{equation}
\label{eq:H_k}
H_k  = \sum_{ll',\sigma} H_{ll',k \sigma} \, b^\dagger_{kl,\sigma}b^{}_{kl',\sigma},
\end{equation}
with matrix elements
\begin{equation}
\label{eq:H_kll'}
H_{ll',k \sigma}=
-\widetilde{t}_{ll',k} + U\delta_{ll'} \langle n_{l, -\sigma}\rangle \,
\end{equation}
that are independent of $m$ due to translational invariance, namely, 
$\langle n_{i \sigma} \rangle \equiv \langle n_{ml,\sigma} \rangle =  
\langle n_{l \sigma}\rangle$. For sites within the same PUC, $\widetilde{t}_{ll',k}$ 
represents the nearest and the next-nearest hopping integrals $t$ and $t^\prime$. For neighboring sites at different PUCs, the hopping terms acquire the phase $e^{\pm i k a}$.

The occupations $\langle n_{l, \sigma}\rangle$ are obtained self-consistently. The 
problem is defined with the help of the eigenenergies $\{\varepsilon_{k\nu,\sigma}\}$ 
and eigenfunctions $\{\varphi_{k\nu,\sigma}(l)\}$ of $H_{ll',k\sigma}$: For a given 
wave number $k$, the $\nu$th state occupation follows the Fermi distribution at zero 
temperature, namely, $\langle n_{k\nu,\sigma}\rangle=\Theta(\mu - \varepsilon_{k\nu,\sigma})$, 
where $\Theta$ is the Heaviside step function and $\mu$ is the chemical potential. 
The probability amplitudes $\varphi_{k\nu,\sigma}(l)$ allow one to calculate the $l$-site occupation for a fixed $k$ using
\begin{equation}
\langle n_{kl,\sigma} \rangle =\sum_{\nu} | \varphi_{k\nu,\sigma}(l)|^2
\langle n_{k\nu ,\sigma} \rangle.
\end{equation}
Finally, the occupation appearing in Eq.~\eqref{eq:H_kll'} is obtained by integrating 
$\langle n_{kl,\sigma} \rangle $ over the Brillouin zone, namely,
\begin{equation}
\label{eq:density}
\langle n_{l,\sigma} \rangle  =\int \! \frac{dk} {{\cal V}_{\rm BZ}}\,\langle n_{kl,\sigma} \rangle 
= \frac{1}{M} \sum_k \langle n_{kl,\sigma} \rangle,
\end{equation}
where ${\cal V}_{\rm BZ}$ is the ``volume" of the Brillouin zone.

The ground-state energy per unit cell $E_0$ is a sum of the occupied self-consistent 
single-particle state energies minus a standard term accounting for double counting 
the on-site Coulomb interaction energy, namely,
\be
E_0 = \sum_{\nu,\sigma}\int \! \frac{dk}{{\cal V}_{\rm BZ}} \,\langle n_{k\nu,\sigma} \rangle 
\ve_{k\nu,\sigma} - \frac{U}{2} \sum_{l,\sigma} \langle n_{l,\sigma} \rangle\langle n_{l,-\sigma} \rangle.
\ee

The $l$th-site magnetization (in units of $\mu_B/2$) is defined as 
\begin{equation}
M_l =  \langle n_{l, \uparrow}\rangle - \langle n_{l, \downarrow}\rangle .
\end{equation}
For chiral nanoribbons, one is frequently interested in the average edge magnetization, 
conveniently defined as
\be
\label{eq:Mchiral}
M = \frac{a_0}{a}\sum_{l\in {\rm edge}} M_l,
\ee
where $a$ is the length of the chiral translation vector and the sum runs over 
the sites of the sublattice ($A$ or $B$) with the largest number of dangling bonds 
along the chiral GNR edge, see Fig.~\ref{fig:chiralGNR}.

The LDOS reads
\begin{equation}
\label{eq:LDOS}
\rho_\sigma(\ve, l) = \sum_\nu \int \frac{dk}{{\cal V}_{\rm BZ}}\, \vert 
\varphi_{k\nu,\sigma}(l)\vert ^2 \delta_\Gamma(\ve - \ve_{k \nu,\sigma}),
\end{equation}
where $\delta_\Gamma$ is the Dirac $\delta$-function broadened over an 
energy range $\Gamma$, taken to be much smaller then the typical  energy 
separation between bands. In turn, the density of states is given by 
${\rm DOS}(\ve)=\sum_{l,\sigma} \rho_{\sigma}(\ve,l)$.

\section{Results}
\label{sec:results}

\subsection{Nanoribbons with zigzag edges}

We begin by presenting the band structure and magnetic properties of zigzag GNRs. 
Part of this material can be  found in the literature (see, for instance, 
Refs.~\onlinecite{Yazyev2010,Wakabayashi2012} for a review), but the analysis
we present serves as an important guide for the subsequent discussion of the chiral GNRs 
results.

\begin{figure}[h!]
\centering
\includegraphics[width=0.95\columnwidth]{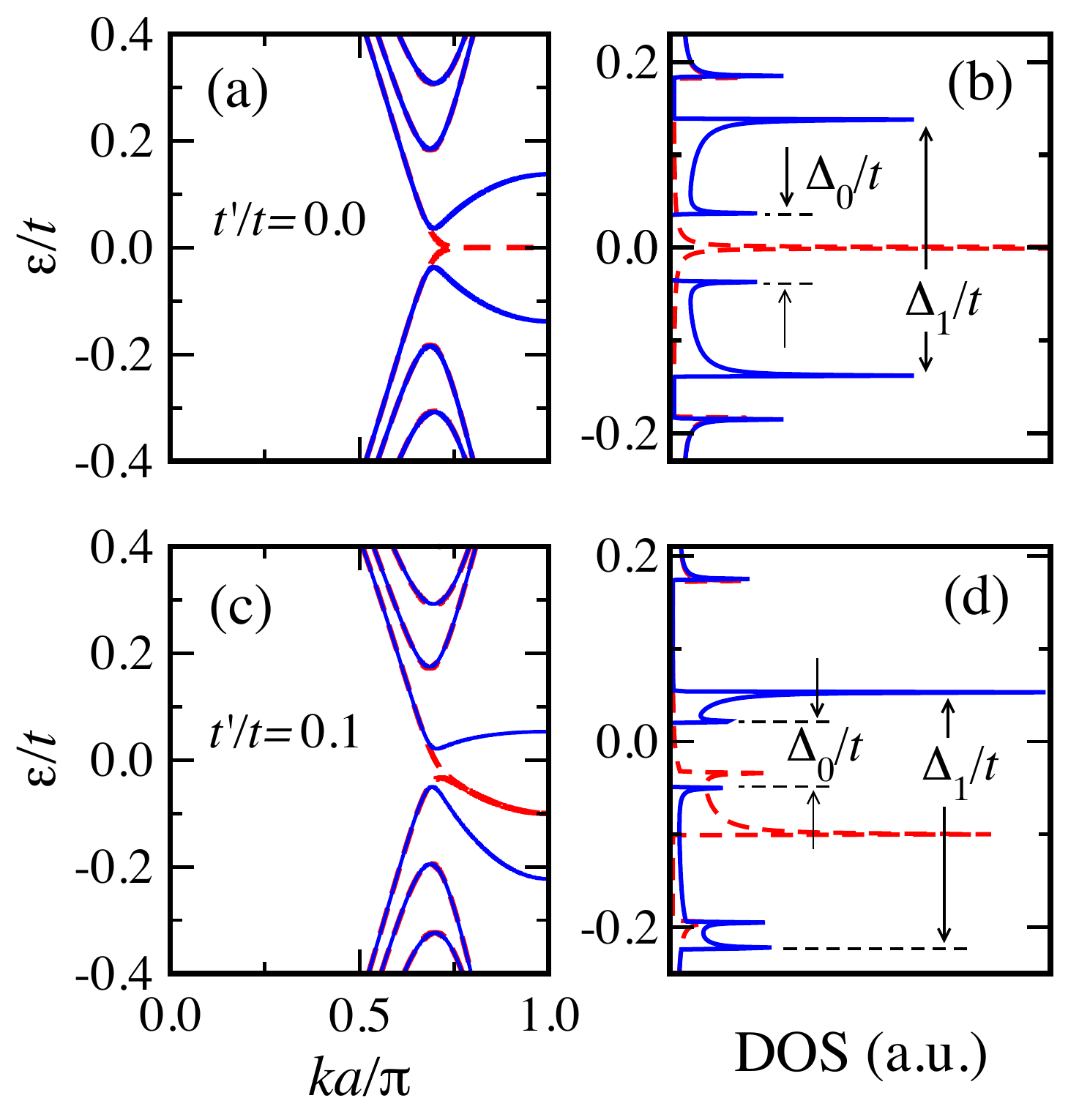}
\caption{(Color online)  Band structure (left column) and corresponding density of states  (right 
column) of a zigzag graphene nanoribbon of $N=24$ for $t'=0$ (upper row) and $t'/t=0.1$ (lower 
row). The solid lines stand for the case of $U/t=1$, while the dashed ones stand for $U=0$. }
\label{fig:zz_bandstructure}
\end{figure}

For $U=0$ and $t'=0$, the dispersionless edge modes enhance dramatically the 
LDOS at the GNR charge neutrality point \cite{Nakada1996}.
When $U\neq0$, due to the Stoner mechanism, the large LDOS at the GNR edges give rise 
to a local magnetization, and the electronic band structure shows a gap around $\ve=0$ for 
$\mu=0$ 
\cite{Nakada1996}. See Fig. \ref{fig:zz_bandstructure}a. 
The ground state shows a parallel spin alignment along each edge and antiferromagnetic
interedge order. This is consistent with Lieb's theorem \cite{Lieb89}, which asserts that 
the ground state of the Hubbard model of a bipartite lattice with nearest-neighbor hopping 
has spin $S=0$.

{\it Ab initio} calculations of zigzag GNR band structures do not show particle-hole symmetry 
\cite{Son2006Nature,Son2006PRL}.
The Hamiltonian \eqref{eq:Hamiltonian} successfully reproduces the DFT band structure 
dispersion relations in the vicinity of the charge neutrality point at the expense of taking 
$t'\neq0$ \cite{Hancock2010}. In this case, Lieb's theorem \cite{Lieb89} is no longer applicable 
and the natural question to ask is how robust is the ground state antiferromagnetic phase. 

This issue is partially understood by a closer analysis of the localized edge states 
as a function of $k$. For $U=0$, the lowest energy $|\ve|$ modes become 
dispersionless at $k\gtrsim 2\pi/3a$, see Fig. \ref{fig:zz_bandstructure}(a). 
As $k$ increases the states become increasingly localized at the GNR edges. 
This behavior persists when $U\neq0$. Accordingly, one spots two characteristic gaps in 
the GNR dispersion relations, see Fig.~\ref{fig:zz_bandstructure}. 
The band gap $\Delta_0$ occurs at the vicinity of the edge 
localization transition. The gap $\Delta_1$ at $k=\pi/a$ is more relevant to the analysis of 
the system magnetic properties. The $k=\pi/a$ point corresponds to the most localized states 
along the GNR edges, which dominate the Stoner magnetization criterion. 

\begin{figure}[h!]
\centering
\centering \includegraphics[width=0.85\columnwidth]{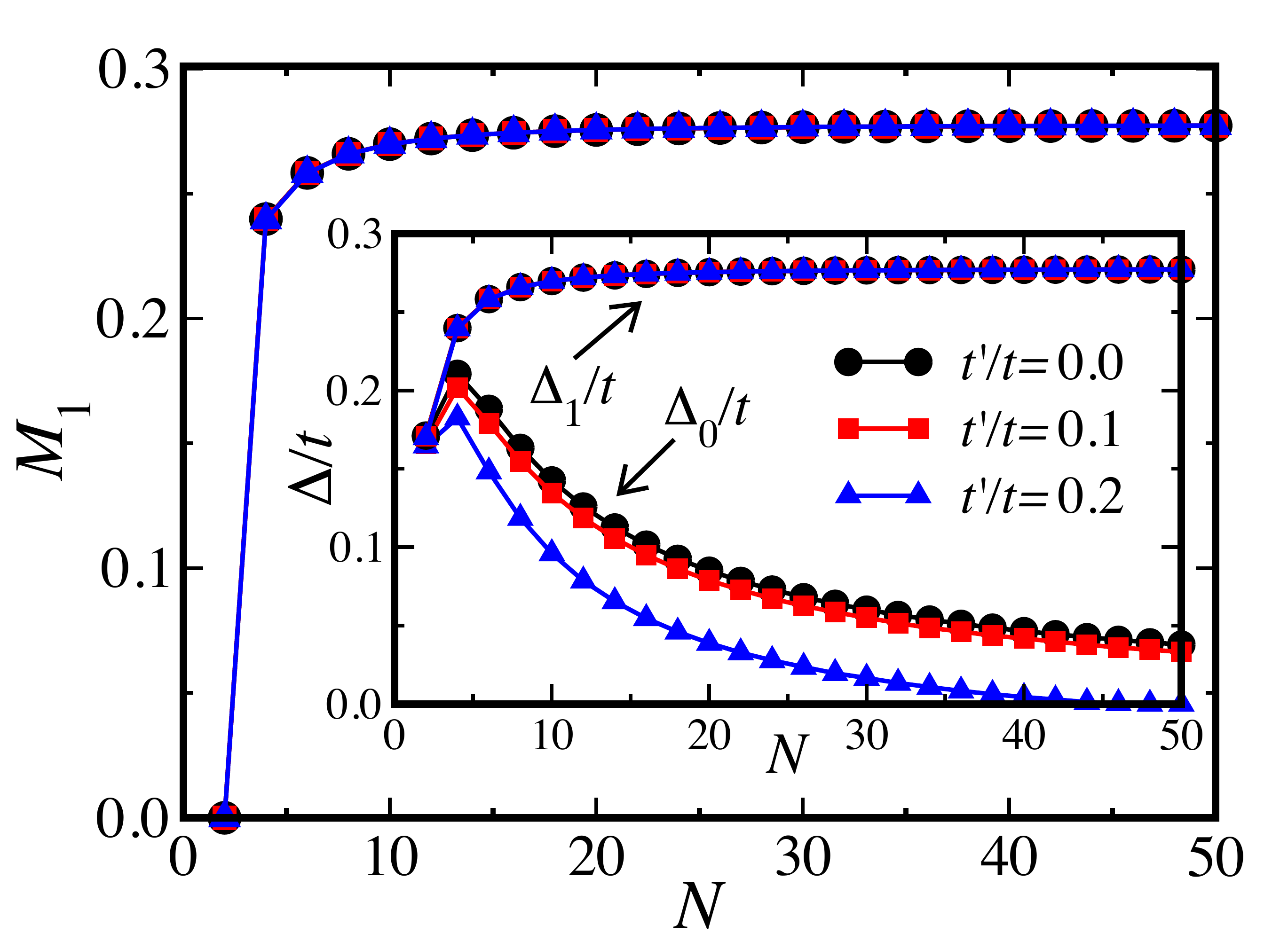}
\vskip0.2cm
\caption{(Color online) Edge magnetization $M_1$ of zigzag graphene nanoribbons 
as a function of their width $N$. Inset: Band gaps $\Delta_0$ and $\Delta_1$ versus $N$. 
In both cases $U/t=1.0$ and $t'/t=0.0,$ $0.1,$ and $0.2$.}
\label{fig:zz_M-Delta}
\end{figure}

Based on this argument, one expects $\Delta_1$ and the edge magnetization $M_1$ to 
be related. This is indeed observed in Fig.~\ref{fig:zz_M-Delta}, that shows $\Delta_0$, 
$\Delta_1$, and $M_1$ as a function of the zigzag GNR width, here conveniently expressed 
in terms of $N$ as the number  of zigzag chains crossing the ribbon transversal direction, 
namely, $W=(\sqrt{3} N/2 +1/\sqrt{3})a_0$. While $\Delta_0$ decreases with increasing 
GNR width, $\Delta_1$ and $M_1$ show a weak $N$ dependence. We stress that neither
$\Delta_1$ nor $M_1$ show a significant dependence on $t'$.

Before proceeding, it is worth noticing that Figs. \ref{fig:zz_bandstructure}(b) and \ref{fig:zz_bandstructure}(d) anticipate some important features we discuss in the 
analysis of the STS spectra of chiral GNRs. While in the particle-hole symmetric 
case $(t'=0),$ the lowest energy $|\varepsilon|$ peaks in the density of states 
can be clearly associated with spin-polarized states, this is not true for $t'\neq0$. 
In Fig. \ref{fig:zz_bandstructure}(d), the peak at $\ve/t\approx-0.19$ corresponds 
to the van Hove singularity of a band unaffected by turning on the interaction $U$, while 
the peak at $\ve/t \approx -0.22$ is the one related to a magnetic state.

\begin{figure}[ht]
\centering \includegraphics[width=0.8\columnwidth]{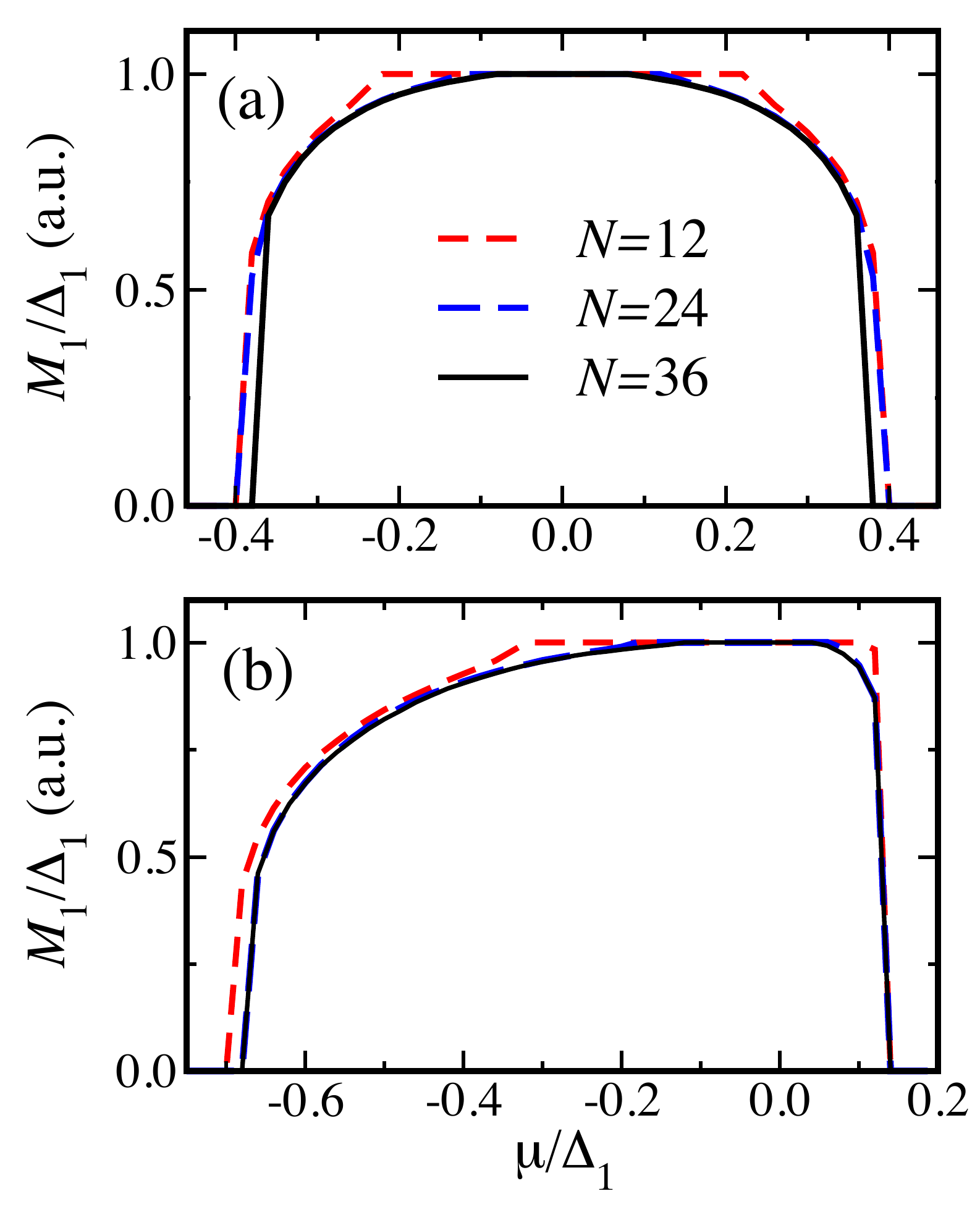}
\vskip-0.2cm
\caption{(Color online) Edge magnetization $M_1$ as a function of the chemical potential $\mu$ 
for zigzag nanoribbons with different GNR widths $N$ for (a) $t'/t=0.0$ and (b) $t'/t=0.1$. 
In both cases $U/t=1$.}
\label{fig:zz_scaled_magnetization}
\end{figure}

Let us now study the behavior of the edge magnetization $M_1$ away from the charge 
neutrality point, or more precisely, as a function of the chemical potential $|\mu|$. 
One expects that, for $\mu$ values close to $\Delta_1/2$, states with opposite spin 
orientation with respect to the ground state start to be occupied and the edge magnetization 
to be suppressed. This is nicely illustrated in Fig.~\ref{fig:zz_scaled_magnetization}, that 
shows that $M_1$ vanishes for $|\mu|\gtrsim\Delta_1/2$. Our calculations also indicate
that $\Delta_1$ slowly decreases with increasing doping (not shown).

Figure \ref{fig:zz_scaled_magnetization} shows that the edge magnetization $M_1$ 
scaled by $\Delta_1$ as a function of the chemical potential $\mu$ shows a  
universal-like behavior for $N\gg 1$.  For $t'=0,$ $M_1/\Delta_1$ is an even function 
of $\mu/\Delta_1$. For very narrow GNRs $(N\lesssim 20),$ $M_1/\Delta_1$ versus 
$\mu$ shows a maximum value, corresponding to a plateaux of a width on the order 
of $\mu/\Delta_0$.  With increasing width, $\Delta_0$ decreases and so does the 
plateaux width. Figure \ref{fig:zz_scaled_magnetization}(a) shows that, already for 
$N\agt 30$, the edge magnetization no longer depends on $N$. The results are qualitatively  
similar for the more realistic case of $t'\neq 0$ \cite{Hancock2010} as illustrated by 
Fig. \ref{fig:zz_scaled_magnetization}(b). 

A lot of attention has been devoted to the study of the competition between the anti- and the ferromagnetic phases in GNRs \cite{Jung2009PRB,Sawada09}. Within the Hubbard 
mean field (for $t'=0$) approximation, Jung and collaborators \cite{Jung2009PRL} studied 
the antiferromagnetic interedge superexchange interaction to estimate the energy difference 
$\Delta E=E_0^{\rm FM}-E_0^{\rm AFM}$ between the antiferromagnetic ground state and 
the energetically lowest ferromagnetic configuration. 
A good fit to numerical calculations is
\be
\Delta E/t = \frac{\alpha}{N^2+ C} 
\ee
where $\alpha=0.245$ and $C=38.9$ for $t'=0$, while $\alpha=0.198$ and $C=45.9$ for $t'/t=0.1$.
These values are smaller than the values reported in Ref.~\onlinecite{Pisani2007} but are in line 
with those of Ref.~\onlinecite{Jung2009PRL} for $t'=0$ and $U=2.0$ eV.
Note, that $\Delta E$ becomes comparable with $k_BT$ at room temperature for 
$N \agt 10$. Hence, for most experimental GNR samples currently available, where $N\gg 1$, 
the interedge interaction is quite negligible.

For doped systems, Lieb's theorem \cite{Lieb89} does not apply, and ground-state phases, 
other than the antiferromagnetic  phases are allowed. Some authors \cite{Jung2009PRB,Sawada09} 
find a very rich phase diagram for zigzag GNRs of $10\leq N\leq30$. We have not performed 
a systematic investigation of non-collinear ground-state solutions as a function of $\mu$ since 
this issue is not central to the goals of our investigation. 
However, it is worth mentioning that, for all tested initial configurations (other 
than the anti- and the ferromagnetic ones), our self-consistent calculations give antiferromagnetic 
ground states for  $|\mu|\lesssim \Delta_1/2$. 
This is an indication that a detailed determination of the magnetic phases can be quite daunting 
and beyond the scope of the model Hamiltonian we use.

What is the ground state configuration for $t'\neq 0$?
In this case, the ferromagnetic ground state at half-filling is no longer forbidden by Lieb's theorem. However, for realistic values of $t'/t$ \cite{Hancock2010}, our calculations only 
lead to (interedge) antiferromagnetic ground states at the charge neutrality point, but we find 
other phases very close in energy. We conclude that, except for very narrow bottom-up grown 
GNRs, the experimental assessment of this phase diagram is very daunting. For this reason, 
we focus our study on the spin alignment along a single edge. 

\subsection{Magnetization in GNR with chiral edges}
\label{sec:chiral}

As discussed above, for zigzag GNRs, the next-nearest-neighbor hopping term modifies 
and reduces the edge LDOS at the charge neutrality point, but surprisingly it does not change
the magnitude of edge magnetization. 
Is the scenario the same for chiral GNRs and how robust is the edge magnetization in this case? 
These are the issues we address in this section.

In the absence of electron-electron interaction, the nearest-neighbor tight-binding model shows 
an enhanced DOS at $\ve=0$ due to a dispersionless band corresponding to edge states. 
Using a continuous rotation of the graphene band structure, it was shown \cite{Akhmerov2008} 
that, in the infinite-width limit, 
\be
\label{eq:rho0}
\rho_0(\theta_c) = \frac{2}{3a_0} \cos\!\left(\theta_c + \frac{\pi}{3}\right),
\ee
where $\rho_0$ is the average edge density of states at $\ve=0$. It is largest for zigzag 
nanoribbons and vanishes for the armchair ones. For chiral GNRs, $0<\theta_c<30^\circ,$
$\rho_0$ shows a nearly linear dependence on $\theta_c$. As discussed previously, the large 
enhancement of $\rho_0$ at the charge neutrality point is key for explaining the edge magnetization 
in GNRs in terms of the Stoner mechanism. As discussed in Ref.~\onlinecite{Yazyev2011}, although 
the edge magnetization $M$ is proportional to $\rho_0$, the band gap $\Delta_0$ (for $t'=0$) is 
related to $U/t$.

The degeneracy of the zero-energy dispersionless modes (for $U=0$) can be understood 
in terms a band folding scheme put forward in Ref.~\onlinecite{Jaskolski2011}: One can 
use Eq.~\eqref{eq:chiral_vector_projection} to write the chiral vector in terms of zigzag and 
armchair projections, namely, $ {\bf C}_h = {\bf C}_{h,{\rm zz} }+ {\bf C}_{h,{\rm ac}} = 
(n-m){\bf a}_1 + m({\bf a}_1+{\bf a}_2)$.
Since the armchair component does not lead to edge states, minimal edge $(n,m)$ GNRs show a 
spectrum similar to the $(n-m,0)$ zigzag ones close to $\ve=0$.
By repeatedly folding the bands of the zigzag edge GNR $(1,0)$, one finds the band structure of a 
$(S, 0)$ edge ribbon. $S$ is conveniently written as $S=I+3P$, where $I=1,2,3$ and $P=0,1,2,\cdots.$
For $I=1$ and $2$, the spectrum has a Dirac-like point at $k\approx 2\pi/3$, while for $I=3,$ the 
Dirac point moves to $k=0$. For $S\ge 3,$ the zero-energy states extend over the whole Brillouin 
zone. The degeneracies of the zero-energy states are either $2P$ or $2(P+1)$ depending on $I$ as 
illustrated in Fig. 2 of Ref.~\onlinecite{Jaskolski2011}. It is simple to show that, in the limit of $S\gg 1$, 
the folding rule leads to the $\rho_0$ given by Eq.~\eqref{eq:rho0}.

\begin{figure}[h]
\centering 
\includegraphics[width=0.95\columnwidth]{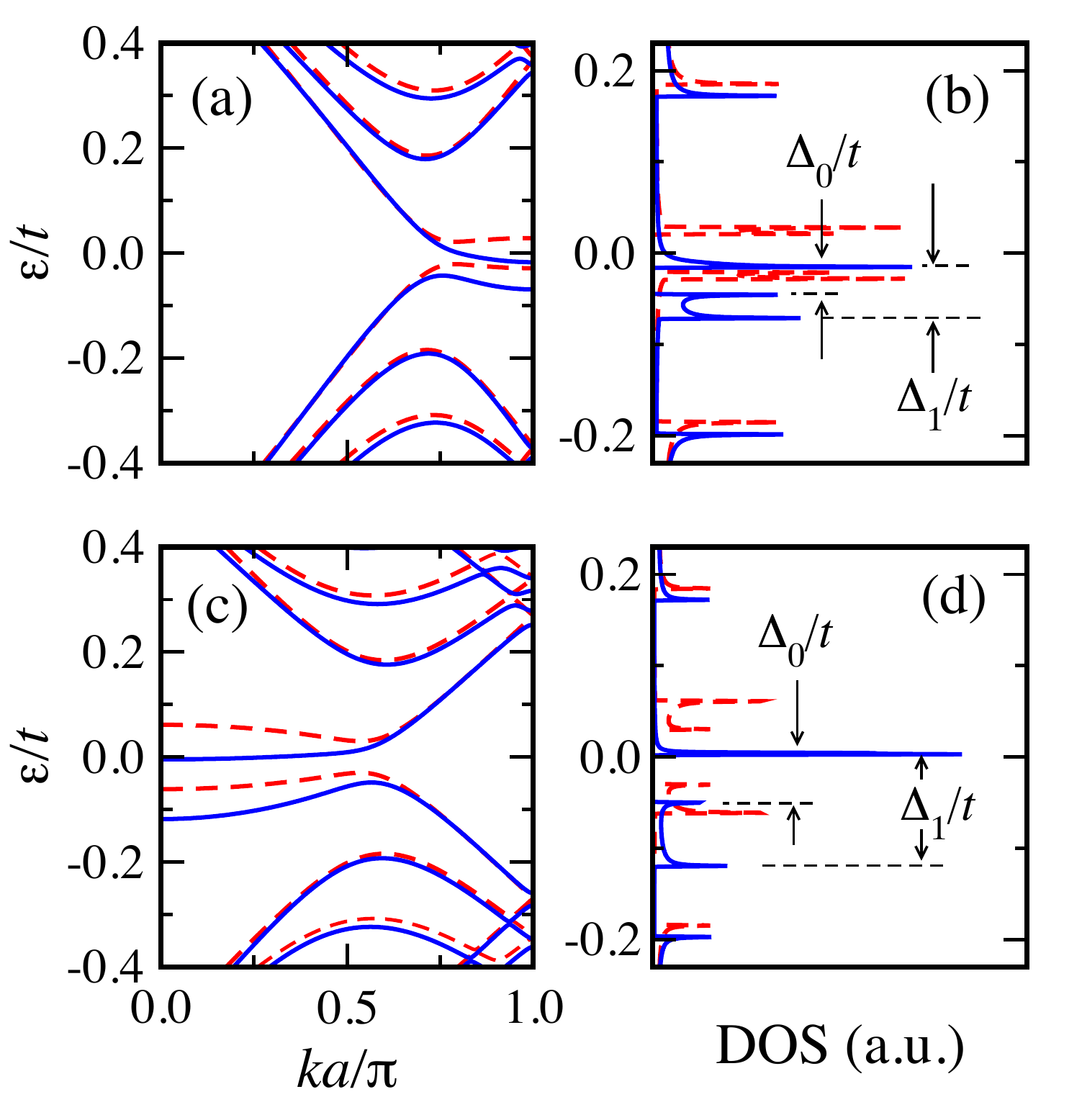} 
\vskip-0.2cm
\caption{(Color online) Electronic band structure for GNRs of width $w=12$ and $U/t=1$, with 
chiralities (a) $(2,1)$ and (c) $(3,1)$. Corresponding density of states for the (b) $(2,1)$ and 
(d) $(3,1)$ chiralities. The solid lines stand for the case of $t'/t=0.1$, while the dashed ones stand
for $t'/t=0.0$. The energy gaps $\Delta_0$ and $\Delta_1$ are only indicated for the $t'/t=0.1$ case. }
\label{fig:chiral_bands(2_1)(3_1)}
\end{figure}

In Fig.~\ref{fig:chiral_bands(2_1)(3_1)}, we show the band structures obtained using the model 
Hamiltonian of Eq.~\eqref{eq:Hamiltonian} for the chiralities  (a) $(2,1)$ with $\theta_c = 19.1^\circ$ 
and (b) $(3,1)$ with $\theta_c = 13.9^\circ$. Here, we take $U/t=1$ and consider the cases of 
$t'=0$ and $t'/t =0.1$, both at half-filling.
As before, $t'\neq0$ breaks the particle-hole symmetry. As a result, the band gap $\Delta_0$ 
goes to zero in most cases, even for very narrow nanoribbons. In distinction to the zigzag case, 
the $k$ point corresponding the maximally localized states at the edges depends on the chirality, 
namely, $ka=\pi$ [Fig. \ref{fig:chiral_bands(2_1)(3_1)}(a)] for the chirality (2,1) and $k=0$ 
[Fig. \ref{fig:chiral_bands(2_1)(3_1)}(c)] for (3,1).

\begin{figure}[ht]
\centering \includegraphics[width=0.9\columnwidth]{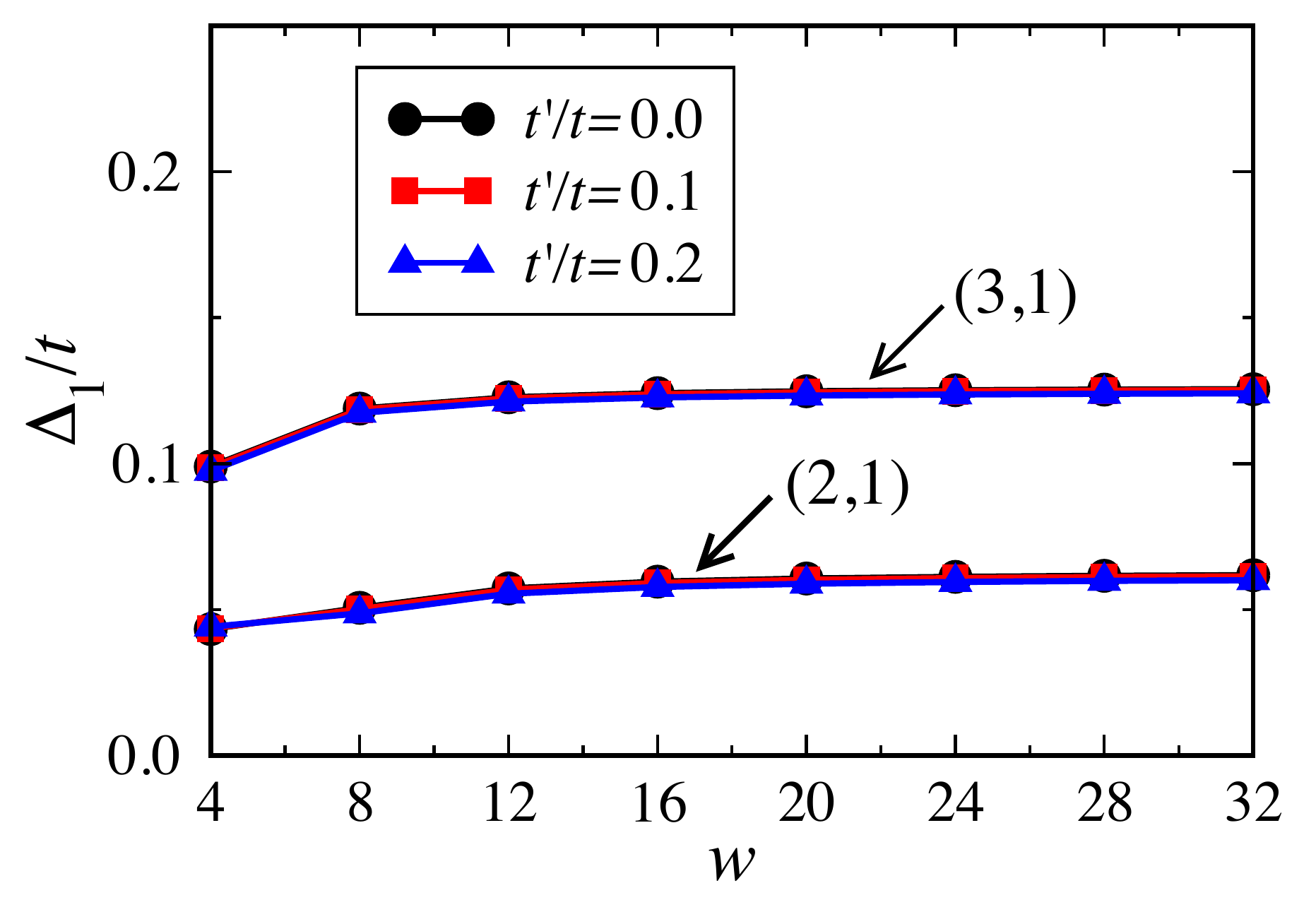}
\caption{(Color online) Gap $\Delta_1$  as a function of the GNR width 
$w$ for different next-nearest-neighbor hopping parameters $t'/t$. Here, we use $U/t=1.0$.}
\vskip-0.2cm
\label{fig:chiral_gap1_vs_w}
\end{figure}

Accordingly, we define $\Delta_1$ as the energy gap around $\ve\approx 0$ calculated at the $k=\pi/a$ point, 
corresponding to maximally localized edge states. Figure \ref{fig:chiral_gap1_vs_w} shows 
$\Delta_1$ as a function of the GNR width $w$ for chiralities ($3,1$) and $(2,1)$.
We find that $\Delta_1$ is independent of $t'$ within the parameter range of that fits the 
DFT calculations. \cite{Hancock2010}  Hence, the numerical results indicate that $t'\neq0$ 
does not significantly change the edge localized states.

Figure \ref{fig:chiral_gap1_vs_w} also shows that $\Delta_1$ increases with $w$ for very 
narrow nanoribbons and becomes almost independent of the GNR width for $w\agt 10$. 
Other chiralities show a similar behavior (not shown here). These observations make it possible 
to relate our findings based on calculations for GNRs of $w\lesssim20$ to experimentally 
realistic sizes, where $w\approx 20 \cdots 50$ \cite{Tao2011}.

\begin{figure}[h!]
\centering \includegraphics[width=0.8\columnwidth]{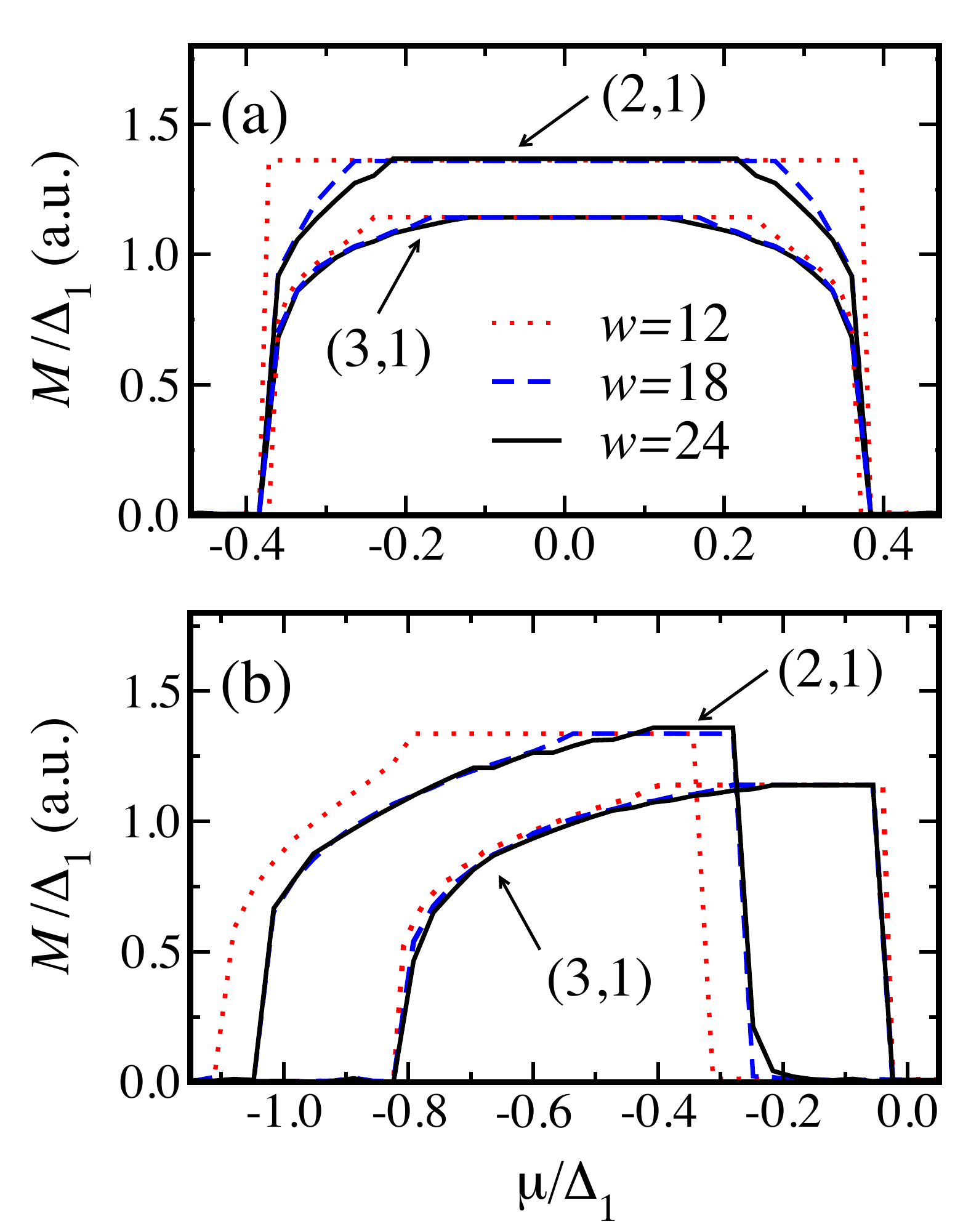}
\vskip-0.2cm
\caption{(Color online) Edge magnetization $M$ as a function of the chemical potential 
$\mu$ for chiral nanoribbons of different widths $w$ for (a) $t'/t=0.0$ and (b) $t'/t=0.1$. 
In both cases $U/t=1$.}
\label{fig:chiral_scaled_magnetization}
\end{figure}

We now turn to the analysis of the edge magnetization as a function of doping (or chemical 
potential $\mu$). Figure \ref{fig:chiral_scaled_magnetization} shows the magnetic moment per 
edge unit length $M$ versus the chemical potential $\mu$, scaled by $\Delta_1$. Here we use this $U/t=1.0$. We find that, for sufficiently wide GNRs ($w \agt 10$), the magnetization $M/\Delta_1$ 
as a function of $\mu/\Delta_1$ becomes independent of $w$. This behavior is 
obtained for both the $t'=0$ and the $t'\neq0$ cases. Figure \ref{fig:chiral_scaled_magnetization} 
indicates  $M$ is not a smooth function of $\mu$. The reason is that the interedge antiferromagnetic 
phase is no longer the ground state of these GNRs away from half filling. The phase diagram is very
rich, but $M$ does not change appreciably. For this reason we did not pursue this line of investigation.

\begin{figure}[h!]
\centering \includegraphics[width=0.8\columnwidth]{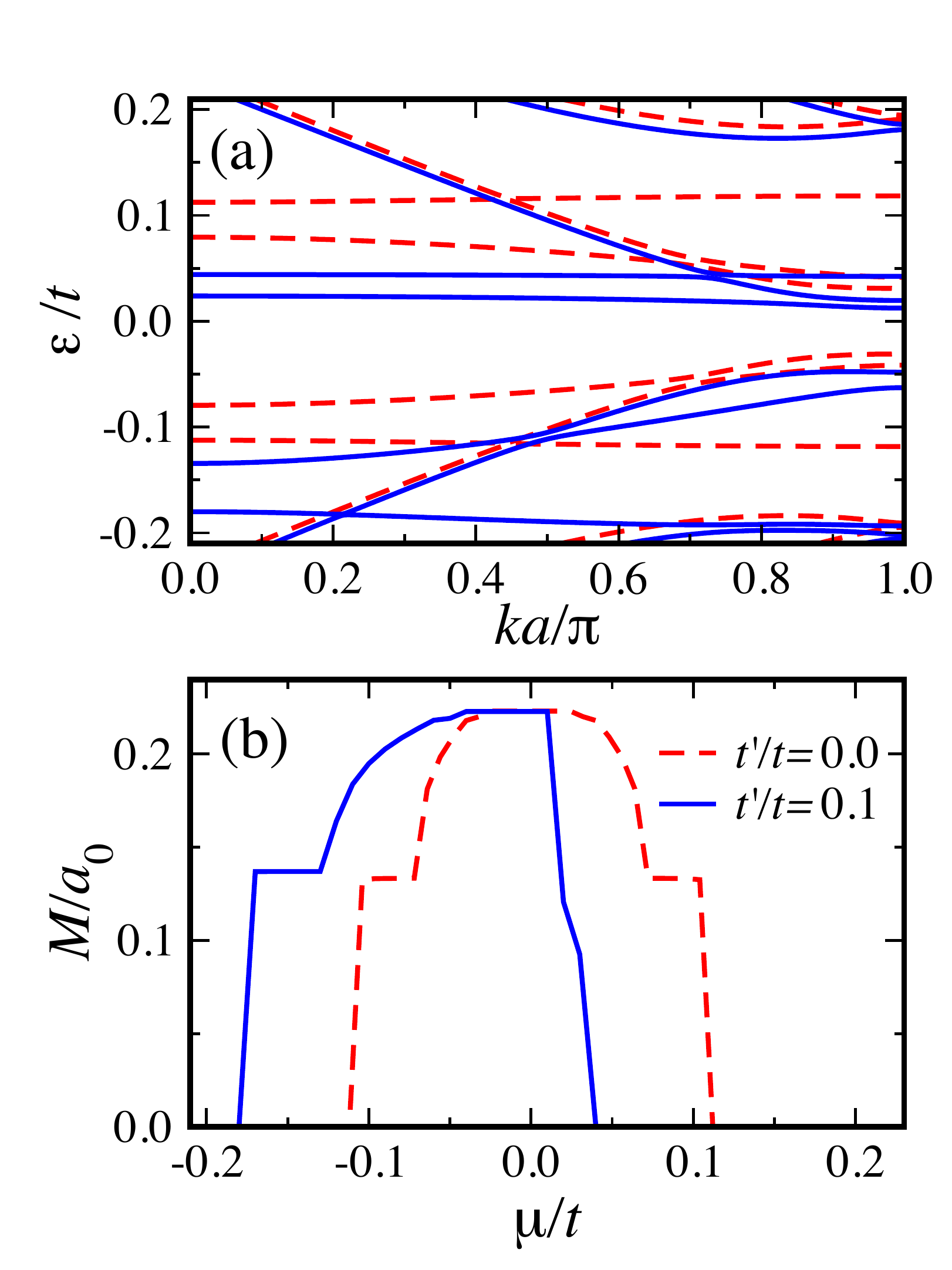}
\caption{(Color online) (a) Band structures of $(8,1)$ chiral graphene nanoribbons 
of $w = 12$ and (b) edge magnetization $M/a_0$ as a function of the chemical potential 
$\mu/t$. The dashed (red) lines stand for the case of $t'/t = 0.0$ and the solid 
(blue) ones stand  for $t/t = 0.1$.}
\vskip-0,3cm
\label{fig:8_1}
\end{figure}		

The chiralities we address above show a strong resemblance to GNRs with zigzag edges. 
We find that by increasing $\theta_c$, for sufficiently wide GNRs,  the edge magnetization $M$ 
decreases almost linearly with $\theta_c$ and, as expected, vanishes for armchair terminations.
The other limit is more interesting. For $U=0$, by decreasing $\theta_c$, one increases $S$ and 
the degeneracy of the zero-energy states modes. For $U\ne 0$, these states split and give rise 
to a complicated band structure around half-filling as illustrated in Fig. \ref{fig:8_1}(a) for the
$(8,1)$ chirality with $S=7, P=2,$ and $I=1$. The corresponding edge magnetization $M$ as 
a function of $\mu$ is shown in Fig.~\ref{fig:8_1}(b).
The latter clearly indicates that the nearly dispersionless modes of  Fig.~\ref{fig:8_1}(a) are the 
ones that contribute most to $M$.

\subsection{LDOS in chiral graphene nanoribbons}
\label{sec:comparison}

As pointed out in the Introduction, the current experimental evidence for edge magnetization 
in GNRs is indirect: The local density of states measured by STS in graphene nanoribbons is claimed to show a behavior consistent with the 
theory for a variety of chiralities \cite{Tao2011,Yazyev2011}.

The STS data main features are the following \cite{Tao2011}: When the tip is placed at 
the GNR edge the measured spectra display two clear peaks close the charge neutrality 
point. As the tip is moved away from the edge, the peak amplitudes are quickly suppressed. 
By moving the tip parallel to the edge, the peak amplitudes show modulations, with a period 
consistent with the size of the translation vector, $a=|{\bf C}_h|$ \cite{Tao2011}. In general,
the peak heights show a large asymmetry that remains unexplained.

The experimental peak spacing has been associated with $\Delta_0$ \cite{Tao2011,Yazyev2011}. 
For that, it is necessary to take $U=0.5t$, a value somewhat smaller than the conventional one 
\cite{Hancock2010}, based on the argument of screening effects due to the metallic 
substrate. It is argued that the opening of an inelastic phonon scattering channel at $|\ve| = 65$ meV 
makes it hard to observe higher energy peaks in the DOS. In what follows, we discuss how the nnn 
hopping term changes this picture.

\begin{figure}[h!]
\centering \includegraphics[width=0.8\columnwidth]{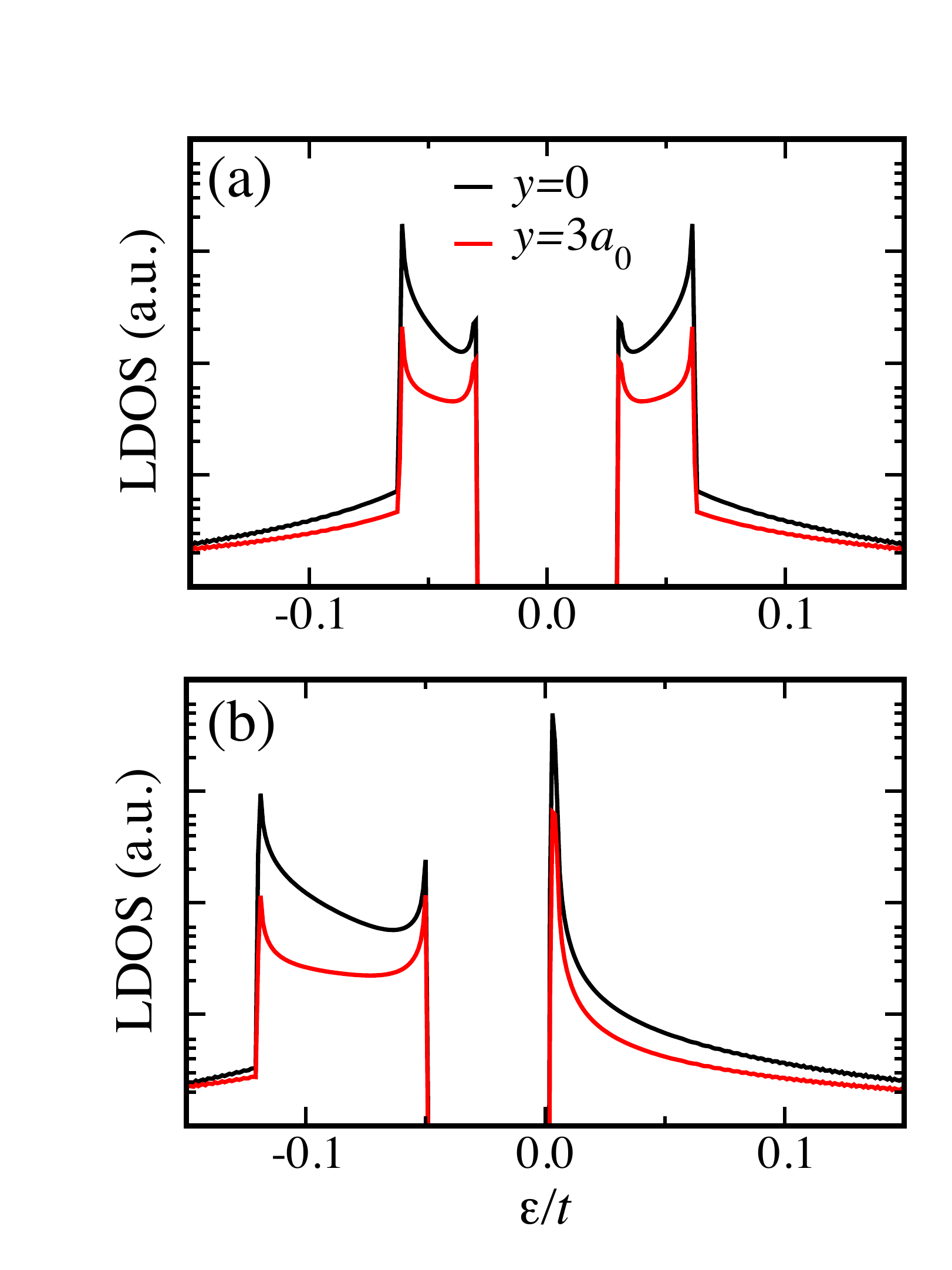}
\vskip-0.2cm
\caption{(Color online) Local density of states for a $(3,1)$ chiral graphene nanoribbon for 
(a) $t'/t=0$ and (b) $t'/t=0.1$.  Here, $w=12$ and $U/t=1$.}
\label{fig:ldos_y_chiral_31}
\end{figure}

In Fig.~\ref{fig:ldos_y_chiral_31} we present the LDOS as a 
function of energy $\varepsilon/t$ for a GNR of chirality $(3,1)$. The LDOS is calculated 
along the edge (referred to as $y=0$) and inside the ribbon along the longitudinal orientation 
($y=3a_0$, in red). Figures \ref{fig:ldos_y_chiral_31}(a) and \ref{fig:ldos_y_chiral_31}(b)
correspond to the $t'/t=0$ and $t'/t=0.1$ cases, respectively. We use $U=t$. The 
LDOS decreases exponentially with increasing $y$, indicating that the peaks in the LDOS
correspond to edge states. 
Note that, for the realistic case of $t'/t=0.1$, the LDOS peak amplitudes become asymmetric 
and the peak spacing can be understood in terms of $\Delta_0$ and $\Delta_1$, defined in
Figs.~\ref{fig:chiral_bands(2_1)(3_1)}(c) and (d).

The GNR chirality $(8,1)$ is experimentally analyzed in detail in Ref.~\onlinecite{Tao2011} [see, 
for example, Fig.~2(c) therein]. Its corresponding low-energy band structure has a more complex one 
than that of the $(3,1)$ case.

\begin{figure}[h!]
\centering \includegraphics[width=0.8\columnwidth]{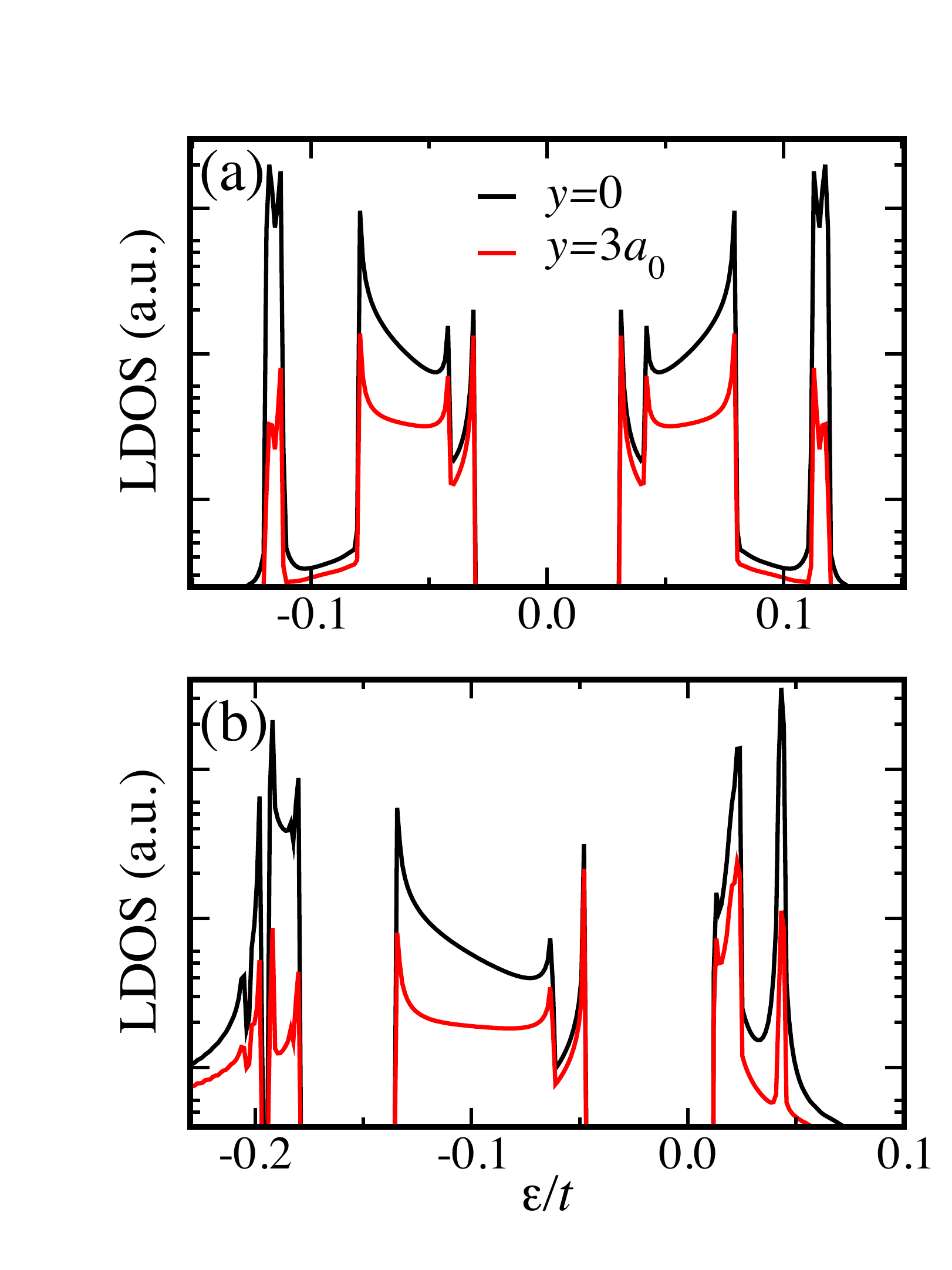}
\caption{(Color online) Local density of states of an $(8,1)$ chiral graphene nanoribbon for 
(a) $t'/t=0$ and (b) $t'/t=0.1$.  Here, $w=12$ and $U/t=1$.}
\label{fig:ldos_y_chiral_81}
\end{figure}

\begin{figure}[h]
\vskip0.3cm
\centering \includegraphics[width=0.9\columnwidth]{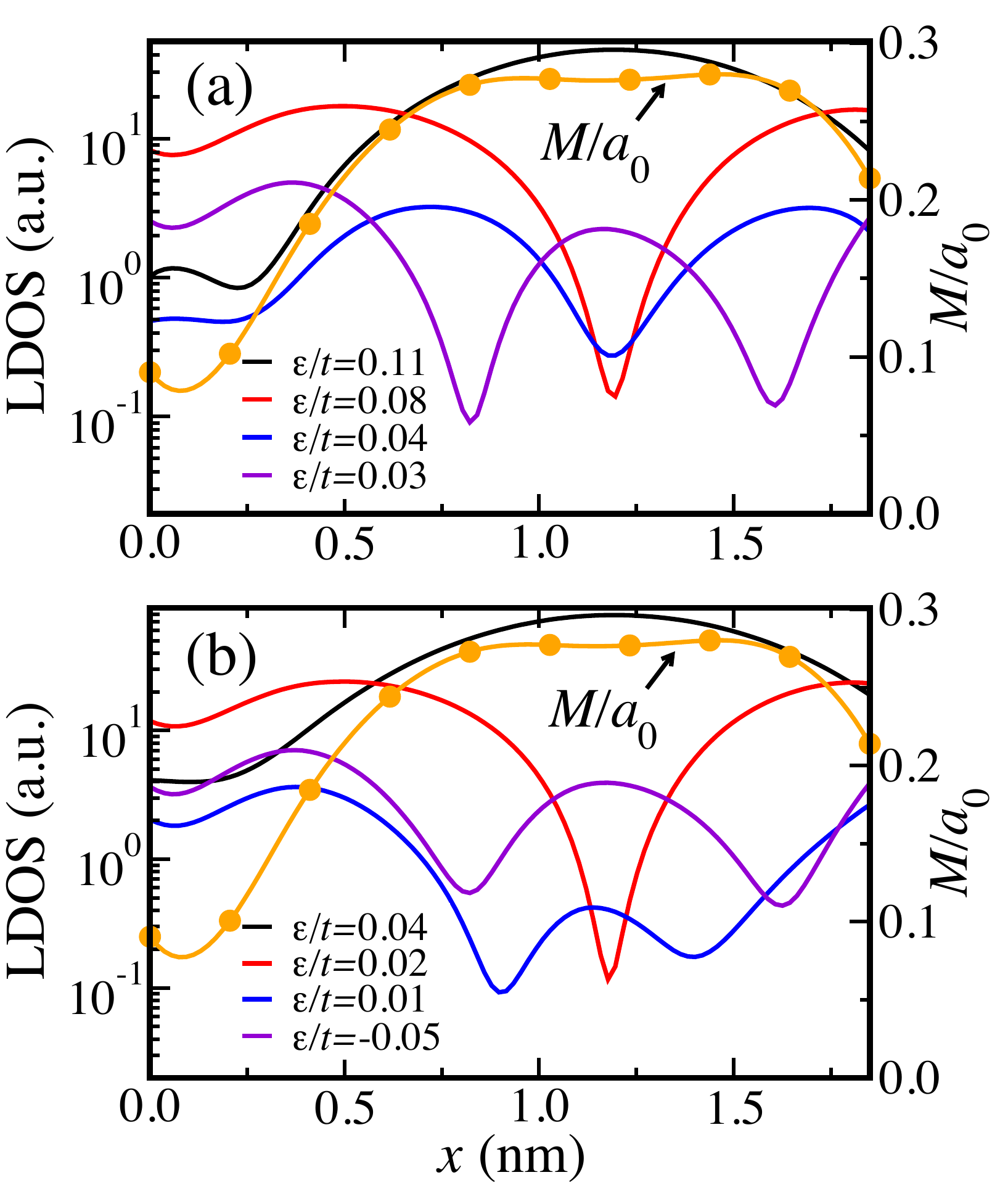}
\caption{(Color online)  Local density of states (logarithmic scale) and edge 
magnetization (linear scale) $M$ of an $(8,1)$ chiral graphene nanoribbon  
for (a) $t'/t=0$ and (b) $t'/t=0.1$. Here, $w=12$ and $U/t=1$.}
\label{fig:ldos_mag_vs_x}
\end{figure}

Figure \ref{fig:ldos_y_chiral_81} shows the LDOS of a GNR with chirality $(8,1)$. From 
Fig.~\ref{fig:8_1} we infer that the states that contribute most to the edge magnetization 
are those corresponding to the dispersionless bands. For $t'=0$, the latter are located 
at $\varepsilon_1^a/t\approx  \pm 0.08$ and  $\varepsilon_1^a/t\approx  \pm 0.12$. Accordingly, 
Fig.~\ref{fig:ldos_y_chiral_81}a shows that the LDOS peaks at the energies $\varepsilon_1^a/t$ 
and $\varepsilon_2^a/t$ are the ones that are most localized at the GNR edges. For the more 
realistic case of $t'/t=0.1$, the states that predominantly drive the magnetization $M$ 
(see Fig.~\ref{fig:8_1}b), are the flat bands at $\varepsilon_1^b/t\approx-0.18, \varepsilon_2^b/t
\approx 0.02$, and $\varepsilon_3^b/t\approx 0.04$. The LDOS peak at  $\varepsilon_4^b/t
\approx-0.05$ corresponds to a van Hove singularity of an ordinary band, that is, a band 
whose states are not localized at the GNR edges. Hence, the band gap $\Delta_0$ involves 
localized and delocalized states.

Let us now examine the local magnetization along the edges. In Fig.~\ref{fig:ldos_mag_vs_x} 
we select values of $\varepsilon/t$ corresponding to the representative sharp peaks of 
Fig.~\ref{fig:ldos_y_chiral_81} and plot the corresponding LDOS and edge magnetization $M$ 
as a function $x$, the position oriented along the GNR edge. The case $t'/t=0$ corresponds to 
Fig.~\ref{fig:ldos_mag_vs_x}(a) and $t'/t=0.1$ corresponds to Fig.~\ref{fig:ldos_mag_vs_x}(b). These figures 
indicate that the edge magnetization $M$ and LDOS corresponding to the dispersionless states 
at $\varepsilon/t=0.11$ (for $t'/t=0$) and $\varepsilon/t=0.04$ (for $t'/t=0.1$) display a very similar 
behavior to $x$. This observation gives further support to the discussion of the previous 
paragraph, corroborating the picture that, for $S\gg1$, the split dispersionless states dominate
the edge magnetization.

\section{Conclusions}
\label{sec:conclusions}

We study the electronic band structure, the local density of states, and the edge magnetization 
of chiral graphene nanoribbons using a $\pi$-orbital Hubbard model in the mean-field approximation. 
We show that the inclusion of a next-nearest hopping term $t'$ in the tight-binding Hamiltonian that 
is necessary for the realistic modeling of the electronic properties of GNRs changes its band structure 
significantly: While $t'\neq0$ has little effect on the average magnitude of the edge magnetization 
at the charge neutrality point, the nnn hopping term largely modifies the behavior of $M$ as a function 
of doping. We believe that these observations call for more realistic analysis of the spin-wave 
excitations in GNRs \cite{Wakabayashi1999,Yazyev2008,Culchac2011}.

The most notable effect of a $t'$ is on the density of states. Our study indicates that the 
interpretation of STM/STS data is very different for $t'=0$ and the more realistic case of $t'/t=0.1$.
In the latter and  for the $(8,1)$ chirality, the energy peak spacing $\delta=\varepsilon_3^b-\varepsilon_2^b$ 
and the peak height asymmetry are consistent with the results reported in Ref.~\onlinecite{Tao2011}. 
However, in analogy to the discussion of $\Delta_1$, we do not expect $\delta$ to depend on the width 
of the GNR $w$. We believe that an experimental LDOS study of the GNR for a fixed chirality and different 
widths can be of great help for the understanding of the edge magnetization in GNRs.

\acknowledgments
We thank R. Capaz and E. Mucciolo for very helpful discussions. 
This work was supported by the Brazilian funding agencies FAPERJ, CAPES, 
CNPq, and INCT - Nanomateriais de Carbono.

%


\end{document}